\documentclass[sigconf,natbib=true]{acmart} 
\AtBeginDocument{%
  }

\copyrightyear{2026}
\acmYear{2026}
\setcopyright{cc}
\setcctype{by-nc-nd}
\acmConference[ICMR '26]{International Conference on Multimedia Retrieval}{June 16--19, 2026}{Amsterdam, Netherlands}
\acmBooktitle{International Conference on Multimedia Retrieval (ICMR '26), June 16--19, 2026, Amsterdam, Netherlands}
\acmDOI{10.1145/3805622.3810780}
\acmISBN{979-8-4007-2617-0/2026/06}

\usepackage{color}
\usepackage{algorithm}
\usepackage{algorithmic}
\usepackage{graphicx}
\usepackage{amsfonts}
\usepackage{multirow}
\usepackage{balance} 
\usepackage{mathrsfs}
\usepackage{makecell}
\usepackage{booktabs,array}
\usepackage{tcolorbox}
\usepackage{enumitem}
\usepackage{amsmath}

\newcommand{\ModelName}{{Rabtriever}}

\begin{document}

\title{Efficient Rationale-based Retrieval: On-policy Distillation from Generative Rerankers based on JEPA}

\author{Teng Chen, Sheng Xu, Feixiang Guo, Xiaoyu Wang, Qingqing Gu, Hongyan Li, Luo Ji$^*$}
\affiliation{%
  \institution{Geely AI Lab} 
  \city{Ningbo} 
  \state{Zhejiang} 
  \country{China}
}

\email{{Teng.Chen2,Sheng.Xu13,Feixiang.Guo1,e-Xiaoyu.Wang1,Qingqing.Gu3,Hongyan.Li10,Luo.Ji1}@Geely.com}

\thanks{*Corresponding author}

\renewcommand{\shortauthors}{Chen et al.}

\begin{abstract}


Unlike traditional fact-based retrieval, rationale-based retrieval typically necessitates cross-encoding of query-document pairs using large language models, incurring substantial computational costs. To address this limitation, we propose Rabtriever, which independently encodes queries and documents, while providing comparable cross query-document comprehension capabilities to rerankers. We start from training a LLM-based generative reranker, which puts the document prior to the query and prompts the LLM to generate the relevance score by log probabilities. We then employ it as the teacher of an on-policy distillation framework, with Rabtriever as the student to reconstruct the teacher's contextual-aware query embedding. To achieve this effect, Rabtriever is first initialized from the teacher, with parameters frozen. The Joint-Embedding Predictive Architecture (JEPA) paradigm is then adopted, which integrates a lightweight, trainable predictor between LLM layers and heads, projecting the query embedding into a new hidden space, with the document embedding as the latent vector. JEPA then minimizes the distribution difference between this projected embedding and the teacher embedding. To strengthen the sampling efficiency of on-policy distillation, we also add an auxiliary loss on the reverse KL of LLM logits, to reshape the student's logit distribution. Rabtriever optimizes the teacher's quadratic complexity on the document length to linear, verified both theoretically and empirically. Experiments show that Rabtriever outperforms different retriever baselines across diverse rationale-based tasks, including empathetic conversations and robotic manipulations, with minor accuracy degradation from the reranker. Rabtriever also generalizes well on traditional retrieval benchmarks such as MS MARCO and BEIR, with comparable performance to the best retriever baseline. 

\end{abstract}



\received{13 February 2026}
\received[revised]{15 April 2026}
\received[accepted]{25 April 2026}


\maketitle

\section{Introduction}

Dense retrieval has emerged as a dominant paradigm of information retrieval (IR), which encodes both queries and documents as dense embeddings within a shared latent space, with relevance typically measured via semantic textual similarity \cite{asai2023tart,su2023instructorXL,wang2024E5}. Nevertheless, the growing adoption of retrieval-augmented generation (RAG) systems increasingly demands more sophisticated forms of knowledge comprehension, requiring retrievers to move beyond the mere matching of explicit facts toward nuanced, often implicit, interpretations of knowledge \cite{gao2024retrievalaugmentedgenerationlargelanguage,zhao2024retrievalaugmentedgenerationaigeneratedcontent,zhao2024retrievalaugmentedgenerationrag,fan2024RAGmeetingLLMs}. This evolution aligns with the knowledge stratification framework proposed by \cite{zhao2024retrievalaugmentedgenerationrag}, which defines four ascending levels of retrieval complexity and depth:\\
\noindent $\bullet$ \textit{Level-1: Explicit Facts}, $\bullet$ \textit{Level-2: Implicit Facts}\\
\noindent $\bullet$ \textit{Level-3: Interpretable Rationales}, $\bullet$ \textit{Level-4: Hidden Rationales}\\
While L1 and L2 correspond to conventional \textbf{fact-based retrieval}, L3 and L4 call for \textbf{rationale-based retrieval}, where the system must comprehend the underlying reasoning, decision logic, or expert knowledge that connects a query to a relevant document. For instance, consider \textit{a software developer who derives actionable coding decisions not only from API documentation alone, but also by examining detailed debugging histories}. In such advanced retrieval scenarios, the relationship between query and document often extends far beyond straightforward semantic or paraphrastic similarity, exposing a fundamental limitation of mainstream dense retrievers \cite{10.1007/978-3-031-88714-7_27}.




Leveraging the extensive domain knowledge and sophisticated reasoning capabilities, LLMs have recently been adopted as backbone architectures for dense retrievers, offering a promising avenue for rationale-based retrieval. LLM-based \textbf{retrievers} typically approach this challenge either through prompting strategies \cite{li2024llama2vec} or integrating with generative objectives \cite{muennighoff2024GritLM,zhang2024onegen}. Nevertheless, they still encode queries and documents independently during inference, lacking a contextualized understanding of their online interactions. In contrast, LLM-based \textbf{rerankers} \cite{muennighoff2022sgpt,sun-etal-2023-chatgpt,ma2024RepLLaMA,10.1007/978-3-031-88714-7_27} jointly encode the concatenated query–document input, thereby enabling deeper and fine-grained reasoning about rationale-based relevance. However, such rerankers require online encoding of every query–document pair, incurring prohibitive computational overheads, which limit their feasibility for dense retrievers in large-scale retrieval systems. Instead, rerankers are generally deployed as the downstream of retrievers in multi-stage text retrieval \cite{ma2024RepLLaMA}.


\begin{figure}[!htbp]

\centering
  \includegraphics[width=0.45\linewidth]{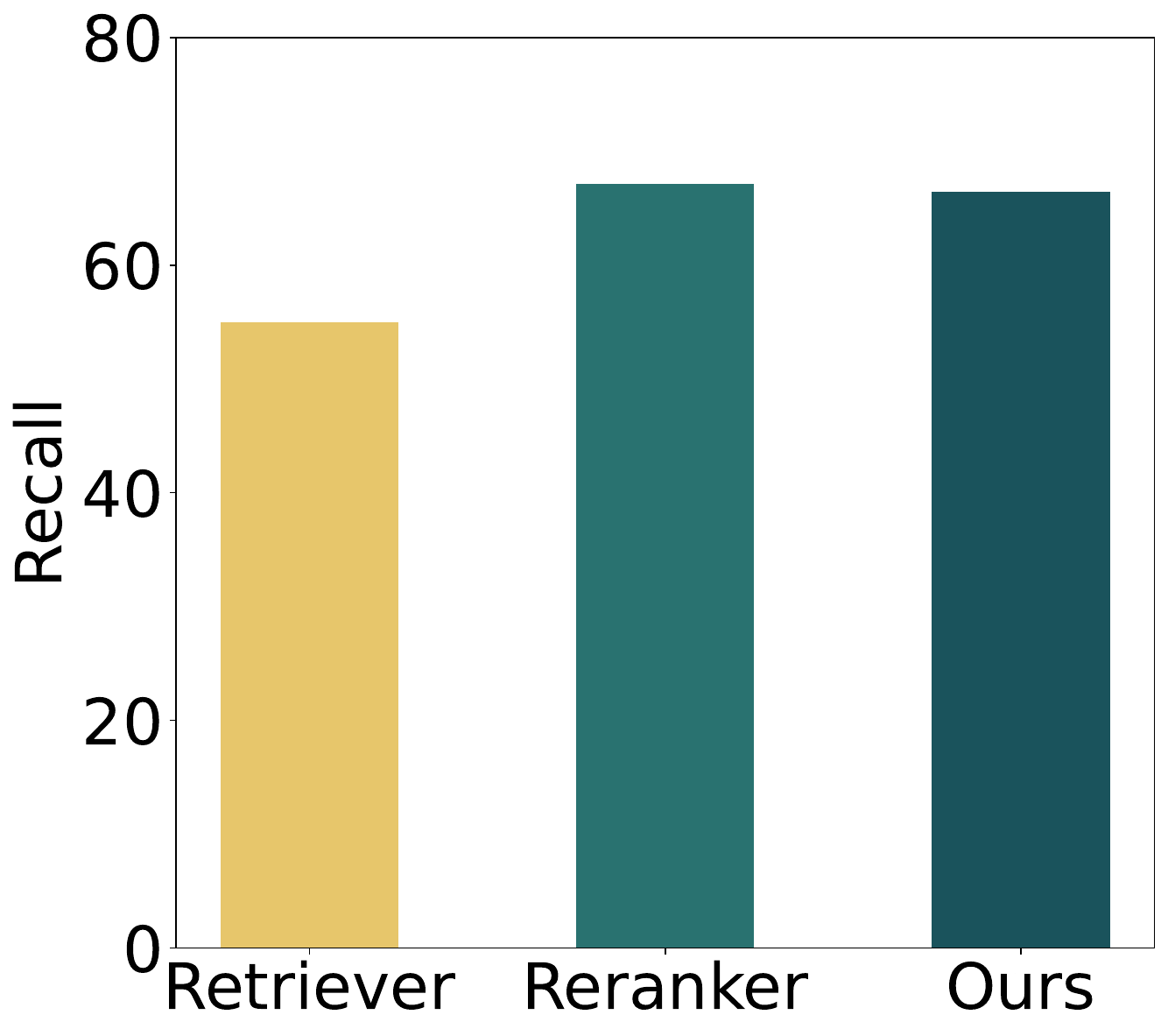}
  \hspace{0.1in}
  \includegraphics[width=0.45\linewidth]{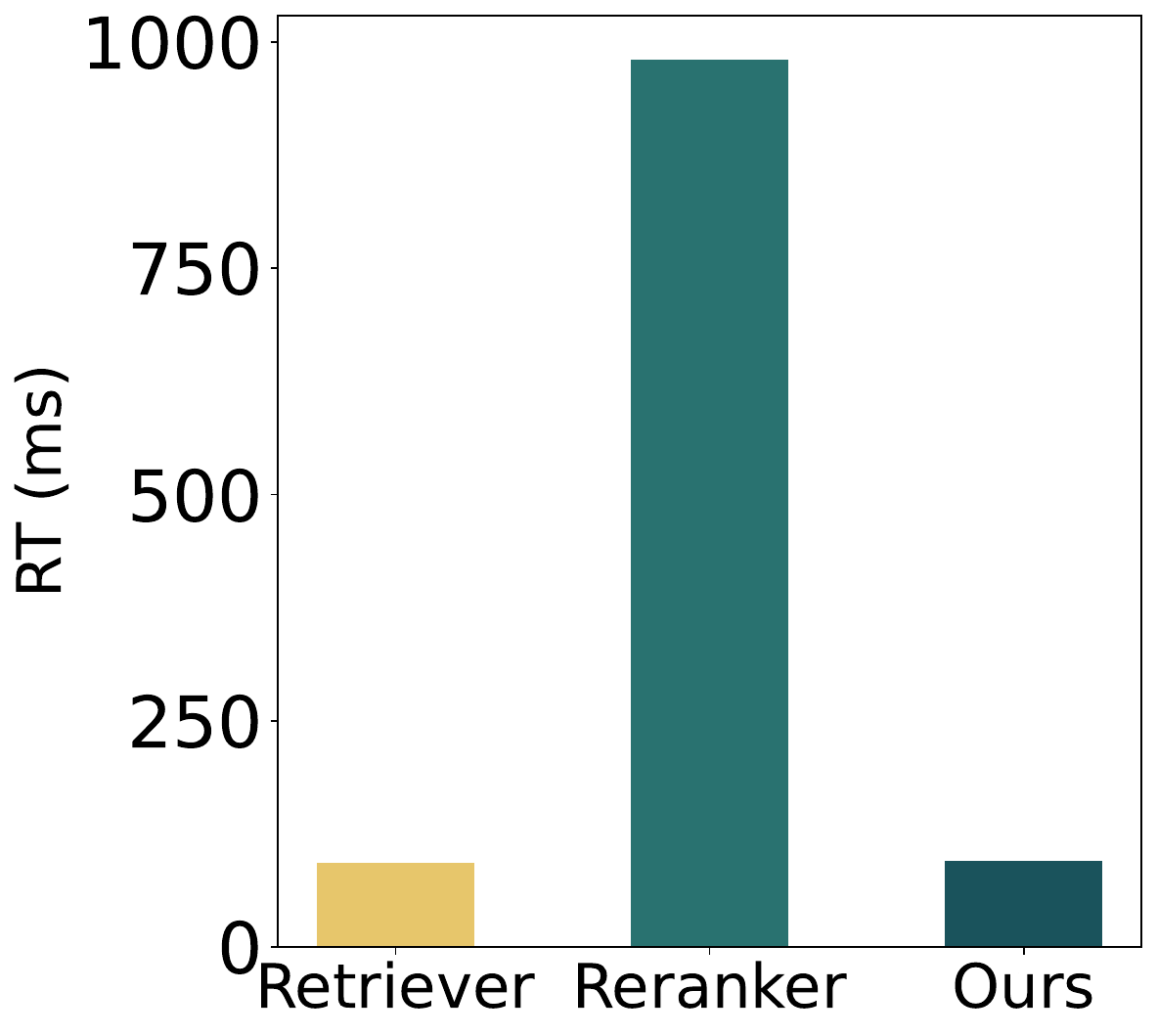}
  \vspace{0.1in}
  \includegraphics[width=0.9\linewidth]{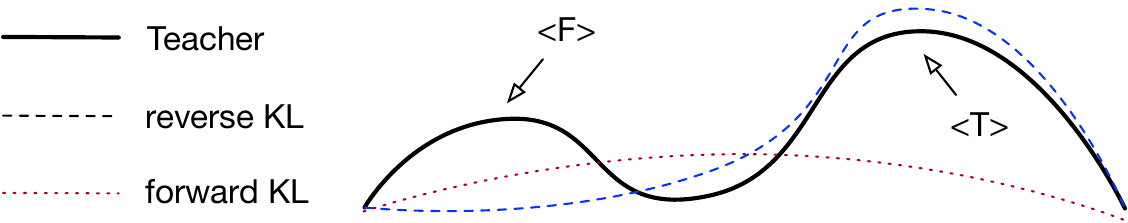} \\
  \caption{Top: By effective distillation, {\ModelName} achieves a similar accuracy to Reranker and a similar speed to Retriever. Bottom: The teacher has two peaks (<T> and <F>) after finetuning; forward KL tends to learn averaged behaviors, while reverse KL helps the student seek the most typical mode \cite{gu2024minillm}.}
  \label{fig:snapshot}
\vspace{-0.1in}
\end{figure}

To resolve this tension, the purpose of this study is to provide a reliable and practical rationale-based retriever, \textbf{with comparable retrieval accuracies to rerankers, and similar inference speeds to traditional retrievers}. The core of our approach lies in an on-policy distillation\footnote{\url{https://thinkingmachines.ai/blog/on-policy-distillation/}} strategy, which transfers cross-comprehension capability of a reranker \textbf{teacher} into a retriever \textbf{student}, reconstructing the reranker's embedding and logit distributions by lightweight cross-computations. To achieve this purpose, we adopt the relevance generation paradigm proposed by \citet{sun-etal-2023-chatgpt}, which puts the document \textit{prior to} the query within the textual input, and produces the relevance score by a generative task. The casual attention mechanism of LLM then naturally ensures the document embeddings are identical between \textbf{teacher} and \textbf{student}, such that they can be pre-computed offline. 





\begin{figure}[!htbp]
\centering
  \includegraphics[width=0.95\linewidth]{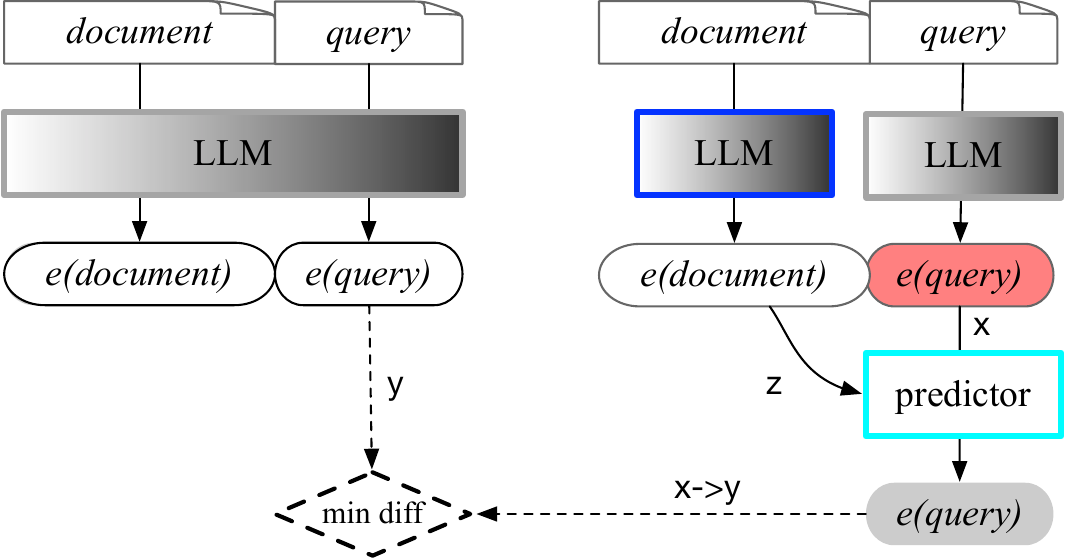}
  \caption {Block diagram visualization of our paradigm. The reranker (left) jointly encodes document and query, with all computations online; {\ModelName} (right) encodes document and query separately, with the query embedding unconditioned on document (marked in \textcolor{red}{red}). We adopt the JEPA paradigm, where a predictor first converts the query embedding ($e_x$) into a new latent space ($e_{x \rightarrow y}$) with the document embedding played as the latent control ($z$), then the difference energy between $e_{x \rightarrow y}$ and the teacher's query embedding ($e_y$) is minimized. The \textcolor{blue}{blue} outline indicates the offline computation; the \textcolor{cyan}{cyan} outline indicates lightweight calculation.}
  \label{fig:paradigm}
\end{figure}

In this paper, we first finetune a \textbf{teacher} based on the LaHoRe framework \cite{10.1007/978-3-031-88714-7_27}, which prompts the LLM to answer "True" or "False" via a binary-choice instruction, and yields the relevance score as the relative log-probability ratio on binary answers. After that, we implement the \textbf{student} with the same LLM backbone, with parameter initiated from the \textbf{teacher} and set frozen. We then leverage a self-supervised framework called Joint-Embedding Predictive Architecture (JEPA) \cite{841b6929fdd44bd58434b6ac370439e2}, in which a trainable \textbf{predictor} is inserted between LLM layers and heads, to project the \textbf{student}'s query embedding into a latent space that is implicitly conditioned on the document embedding. We implement the predictor with a 2-layer MLP and an elementary-multiplication operator to ensure it is sufficiently lightweight. Finally, we adopt both feature-based and logit-based paradigms of white-box knowledge distillation \cite{mansourian2025a}, where the first corresponds to the energy-based modeling (EBM) framework of JEPA, by minimizing MSE between \textbf{teacher} and \textbf{student} embeddings; and the second is an auxiliary loss term to reduce the reverse KL between \textbf{teacher} and \textbf{student} logits, which ensures the student focus on target tokens (\textit{i.e.}, "True" or "False") instead of spreading across suboptimal vocabularies \cite{gu2024minillm}  (Figure \ref{fig:snapshot} Bottom). On the inference stage, we reconstruct the distribution of the \textbf{teacher}'s context-aware query embedding by inferring the predictor, and compute the final relevance score. Figure \ref{fig:paradigm} visualizes this paradigm, and we name it by \textbf{Ra}tionale-\textbf{B}ased Re\textbf{triever}, or \textbf{\ModelName} for abbreviation. 

Following \citet{10.1007/978-3-031-88714-7_27}, we first conduct our experiments on two challenging rationale-based retrieval scenarios:

\noindent (1) A conversational agent identifying the optimal dialogue strategy from an annotated pool to enhance interaction quality, based on specific conversational goals (\textit{e.g.}, emotional support) \cite{liu2021ESconv,sun-etal-2021-psyqa}.\\ 
\noindent (2) A language-grounded robot selecting the most appropriate skill to accomplish a high-level instruction, requiring an understanding of procedural intent and environmental constraints \cite{pmlr-v205-ichter23a,huang2022inner,knowno2023}.
The results substantiate that {\ModelName} matches the retrieval accuracy of state-of-the-art (SoTA) rerankers while achieving orders-of-magnitude reduction in inference latency, as highlighted by Figure \ref{fig:snapshot} (Top). To further verify {\ModelName}'s generalization across standardized large-scale retrieval benchmarks, we then evaluate it by both in-domain retrieval on MS MARCO passage ranking, as well as zero-shot retrieval on the BEIR benchmark. {\ModelName} performs similarly to the most powerful retriever competitor, indicating that our framework generalizes to traditional retrieval tasks. In summary, the principal contributions of this work are fourfold:


\begin{table*}[t]
    \caption{
        Comparison of different retriever architectures, including Retriever, Reranker, and our {\ModelName}. {\ModelName} first initializes from a pretrained reranker, then inserts a trainable latent projector $\mathcal{P}$. Instead of predicting the retrieval scores ($\mathcal{L}$) directly, {\ModelName} employs the EBM paradigm which aims to reconstruct the latent distribution of the cross-encoding reranker. Modules in \textcolor{blue}{blue} indicate the offline computation; modules in \textcolor{cyan}{cyan} indicate lightweight calculation.
    }\label{tab:overview}
    \centering \footnotesize
    \begin{tabular}{llll}
        \toprule
            Definition & \multicolumn{3}{l}{query $D$, document $D$, embedding $e$, relevance score $s$ (ground truth) and $\hat{s}$ (predicted); encoder $\mathcal{M}$, and the (lightweight) score calculator $\mathcal{H}$.} \\
        \midrule
        \midrule
            & Bi-encoder \textbf{Retriever} & Cross-encoder \textbf{Reranker} & \textbf{\ModelName} \\
        \midrule
            Initialize & $\mathcal{M}^Q, \mathcal{M}^D \leftarrow \text{LLM}$ & $\mathcal{M}^{\text{ce}} \leftarrow \text{LLM}$ & $\mathcal{M} \leftarrow \mathcal{M}^{\text{ce}}, \mathcal{H} \leftarrow \mathcal{H}^{\text{ce}}$ \\
        \cmidrule{1-4}
           Forward  &
            $\hat{s}^{\text{be}} \sim \textcolor{cyan}{\mathcal{H}^{\text{be}}(s|e_Q, e_D)} \mathcal{M}^Q(e_Q | Q) \textcolor{blue}{\mathcal{M}^D(e_D | D)}$
            & $\hat{s}^{\text{ce}} \sim \textcolor{cyan}{\mathcal{H}^{\text{ce}}(s|e^{\text{ce}}_{QD})} \mathcal{M}^{\text{ce}}(e^{\text{ce}}_{QD} | Q, D)$ 
            & $\hat{s} \sim \textcolor{cyan}{\mathcal{H}(\hat{s}|e_{QD})} \textcolor{cyan}{\mathcal{P}(e_{QD} | e_D, e_Q)} \mathcal{M}(e_Q | Q) \textcolor{blue}{\mathcal{M}(e_D | D)}$ \\
        \cmidrule{1-4}
            \multirow{1}{*}{Objective} 
            & \multirow{1}{*}{$\min_{\mathcal{\mathcal{M}^Q, \mathcal{M}^D}}{\mathcal{L}(\hat{s}^{\text{be}}, s)}$} 
            & \multirow{1}{*}{$\min_{\mathcal{\mathcal{M}^{\text{ce}}, \mathcal{H}^{\text{ce}}}}{\mathcal{L}(\hat{s}^{\text{ce}}, s)}$} 
            & $\min_{\mathcal{P}}{\mathcal{F}(e_{QD}, e^{\text{ce}}_{QD}) + \text{KL}(\hat{s} || \hat{s}^{\text{ce}})}$, $\mathcal{F}(x, y) \rightarrow 0 \text{ when } x \rightarrow y$ \\
        \bottomrule
    \end{tabular}
\end{table*}

\noindent (1) We propose an efficient LLM-based retriever, which encodes queries and documents independently to ensure efficiency, while performing similarly to rerankers on rationale-based retrieval.\\ 
\noindent (2) We design a distill paradigm on a generative reranker teacher, and employ JEPA to reconstruct its document-conditioned query embedding, by minimizing the latent distribution discrepancy.\\ 
\noindent (3) We conduct comprehensive evaluations across diverse scenarios, including rationale-based tasks (empathetic conversation and language-grounded robotics) and two general retrieval benchmarks.\\ 
\noindent (4) We provide a formal analysis for time complexities of our {\ModelName} and standard rerankers, quantitatively verifying the achieved computational efficiency gains.

\section{Related Works}
\label{sec:related_works}

\paragraph{LLM-based dense retrievers.} Prevail LLM-based retrievers employ the LLM layers to encode query and document separately, and calculate the relevance score by embedding similarity. For example, RepLLama \cite{ma2024RepLLaMA} gets the last-token embedding of Llama and finetunes with the infoNCE loss. E5 \cite{wang2024E5} further improves the performance by augmenting the training sets with diverse, multilingual synthetic datasets. Promptriever \cite{weller2024promptrieverinstructiontrainedretrieversprompted} adds the prompt mechanism on RepLLama with performance improvements on instructed retrieval tasks. Recently, hybrid retrievers have been proposed, with a unified architecture on both representative and generative tasks \cite{muennighoff2024GritLM,zhang2024onegen}. However, these retrievers are primarily designed for fact-based retrieval, which significantly differs from rationale-based retrieval, the main focus of this paper. Although Llama2Vec \cite{li2024llama2vec} tries to bridge the gap by incorporating a simple prompt ("NEXT: ") into the query encoding, explicit comprehension on the cross semantics between query and document is still missing.


\paragraph{LLM-based rerankers.} Within the paradigm of multi-stage text retrievals, the reranker provides a more accurate ranking by jointly encoding the query-document pair, behaving the downstream of the retriever. Its ranking score might be obtained either from the log probability, such as SGPT \cite{muennighoff2022sgpt} and rankGPT \cite{sun-etal-2023-chatgpt}, or by a linear projection like rankLlama \cite{ma2024RepLLaMA}. In this paper, we implement a generative reranker similar to LaHoRe \cite{10.1007/978-3-031-88714-7_27}, which uses a relevance generation paradigm similar to \cite{sun-etal-2023-chatgpt}, while conducting the finetuning as in \cite{ma2024RepLLaMA}. With SoTA performances obtained on several rationale-based retrieval benchmarks, this reranker is then employed as the \textbf{teacher} for further distillation to the efficient retriever, so that the inference speed can be optimized dramatically.

\paragraph{Knowledge transfer from reranker to retriever.} In-batch KD \cite{lin-etal-2021-batch} distills from a bi-encoder teacher; while RocketQAv2 \cite{ren-etal-2021-rocketqav2} and AR2 \cite{zhang2022adversarial} conduct joint training of bi-encoder retriever and cross-encoder reranker. In the contrast, in this paper we distill an LLM-based reranker into a bi-encoder retriever, such that the retrieval performance and efficiency can be balanced. Inspired by MiniLLM \cite{gu2024minillm}, we implement the loss with a reverse KL term, instead of the forward KL, to better shape the student's behavior distributions, effectively leveraging the LLM knowledge on the rationale-based retrieval.



\section{Preliminaries}
\label{sec:preliminary}

\subsection{Joint-Embedding Predictive Architecture}

Given an encoded input-output pair $(x, y)$, a joint-embedding predictive architecture (JEPA) \citep{841b6929fdd44bd58434b6ac370439e2} first converts $x$ into the latent space of $y$ by a predictor $\mathcal{P}(\cdot): \mathbb{R}^{x} \rightarrow \mathbb{R}^{y}$, then minimizes its distributional discrepancy with $y$:
\begin{align}
\min \mathcal{F}(\mathcal{P}_z(x), y) \label{eq:jepa_loss}
\end{align}
where $\mathcal{F}(\cdot, \cdot)$ is nonnegative and can be viewed as the energy between $x$ and $y$, \textit{i.e.}, $\mathcal{F}(x, y) \rightarrow 0$ when $x \rightarrow y$. Subsequently, the JEPA framework belongs to the energy-based model (EBM). $\mathcal{P}$ can be controlled by a prior, non-trainable latent vector $z$.
%

\subsection{Problem Formulation}
\label{sec:problem_formulation}

Given a pool of $n$ document candidates $\{ D \}$ and a user query $Q$, the retriever is required to calculate document relevant scores, $\{ s_i(Q, D_i) \}, 1 \leq i \leq n$, then returns the Top-$k$ $\{ D_i \}$ with respect to the ranking of $s_i$ ($k \leq n$). According to the annotations in the training set, the ground-truth $s_i$ can be labeled by either positive (1) or negative (0). To simplify the computational analysis, we assume the average length of $Q$ and $D$ can be represented by $L_Q$ and $L_D$. 

In the following subsections, we will revisit the paradigms of typical retrievers and rerankers, with corresponding time complexity analysis. We then propose the formulation of {\ModelName}, with a probabilistic view of training objective. Table \ref{tab:overview} summarizes three paradigms. In the table (and also the following formulations), we use \textcolor{blue}{blue} to denote modules which can be calculated off-the-shelf (offline); and use \textcolor{cyan}{cyan} to denote lightweight calculations.

\begin{figure*}[htbp!]
  \includegraphics[width=1\linewidth]{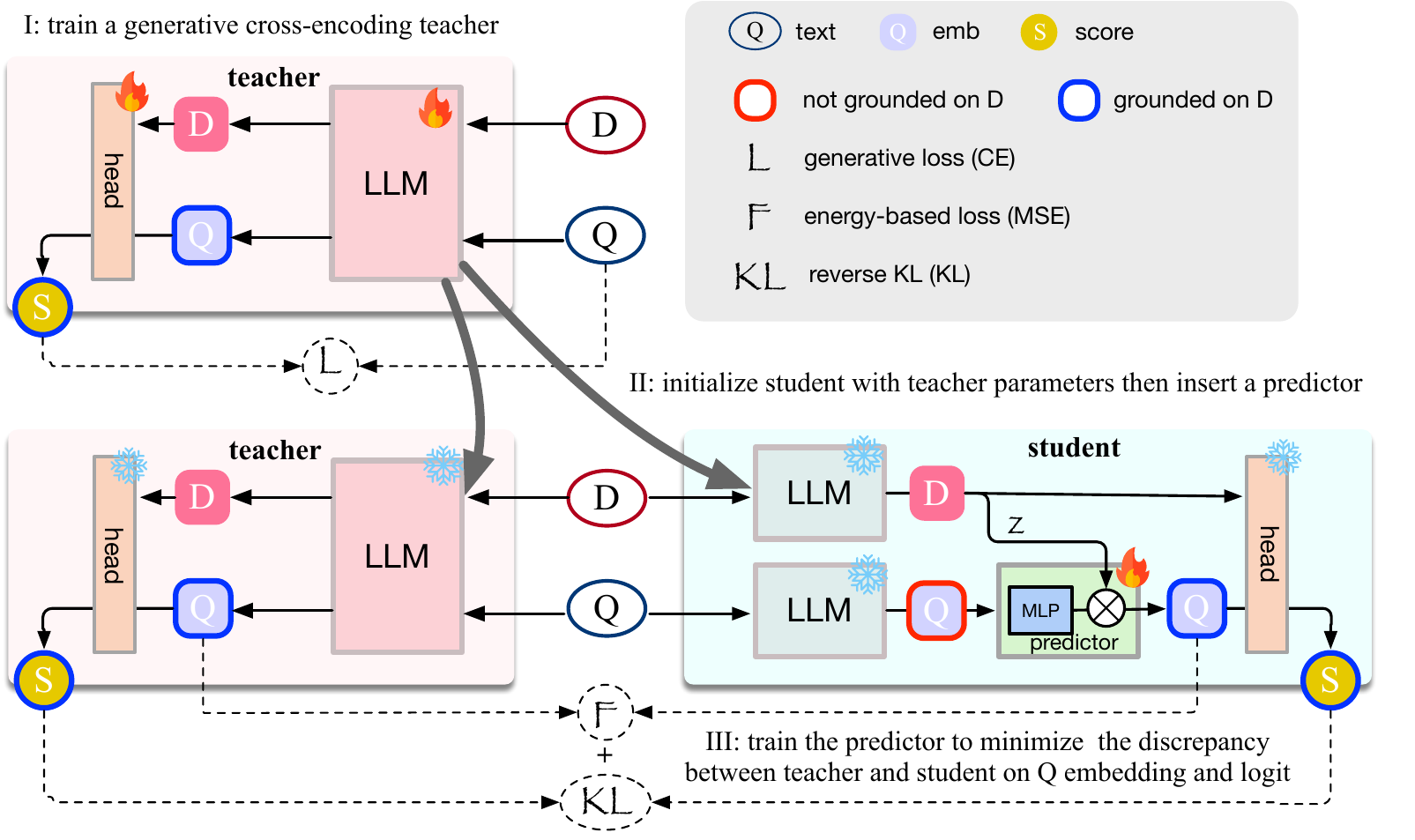} 
  \caption {Pipeline to develop {\ModelName}. I: A \textbf{teacher} cross-encodes query ($Q$) and document ($D$), then generates the retrieval score ($s$). We train it by finetuning an LLM on question-answering samples using the cross-entropy (CE) loss ($\mathcal{L}$). II: We initialize the \textbf{student} from \textbf{teacher}, with parameters frozen. A predictor is then inserted between LLM layers and heads, consisting of an MLP and a multiplication operator. III: The predictor converts the $Q$ embedding into a new embedding, with the $D$ embedding as the latent control variable ($z$). The predicted $Q$ embedding is then fed into the head, which produces the relevance score for \textbf{student}. Both embedding and scoring differences between \textbf{teacher} and \textbf{student} are minimized ($\mathcal{F}$).
  }
  \label{fig:framework}
\end{figure*}

\subsection{Bi-encoder Retriever}

Traditional efficient retrievers typically employ a bi-encoder architecture, which encodes $D$ and $Q$ separately: 
\begin{align}
    \textcolor{blue}{e_D = \mathcal{M}^D(D)} \in \mathcal{R}^d, e_Q = \mathcal{M}^Q(Q) \in \mathcal{R}^d, \textcolor{cyan}{\hat{s}^{\text{be}} = <e_D,  e_Q >} \label{eq:bi_encoder} 
\end{align}
where $\mathcal{M}^Q$ and $\mathcal{M}^D$ are the query and document encoders, while $e_D$ and $e_Q$ are the resulted query and document embeddings, in the dimension of $d$; \textcolor{cyan}{$\textless\cdot,  \cdot \textgreater$} is a lightweight similarity function; The hat on top of $s$ indicates the prediction, and `be' denotes bi-encoder.

\paragraph{Time complexity.} Encoding of $D$ can be pre-computed off-the-shelf (offline). Assume the encoder is has a quadratic complexity (\textit{e.g.}, transformer), then the total time complexity of Eq. (\ref{eq:bi_encoder}) is
\begin{align}
    \underbrace{\textcolor{blue}{n \text{O}(L_D)^2}}_{\text{Offline}} + \underbrace{\text{O}(L_Q)^2 + \textcolor{cyan}{n \text{O}(1)}}_{\text{Online}} \label{eq:time_bi_encoder}
\end{align}

\subsection{Cross-encoder Reranker}

Although efficient, the bi-encoder framework lacks the joint comprehension of $Q$ and $D$, therefore struggles to retrieve a rationale-based query reasonably \cite{10.1007/978-3-031-88714-7_27}. Instead, a cross-encoder can jointly study the semantic or logic correlation between $Q$ and $D$: 
\begin{align}
    e^{\text{ce}}_{QD} = \mathcal{M}^{\text{ce}}(Q, D) \in \mathcal{R}^d, \textcolor{cyan}{\hat{s}^{\text{ce}} = \mathcal{H}^{\text{ce}}(e^{\text{ce}}_{QD}}) \label{eq:cross_encoder}
\end{align}

\paragraph{Time complexity.} In practice, $\mathcal{M}^{\text{ce}}$ can also be implemented with transformers (BERT, GPT, etc), with also quadratic time complexity. However, it must conduct all the computations online:
\begin{align}
    n \text{O}(L_D + L_Q)^2 \quad (\text{Online}) \label{eq:time_cross_encoder}
\end{align}

Since $L_D$, $L_Q$, and $n$ are larger than 1, it is obvious that Eq. (\ref{eq:time_cross_encoder}) is larger than Eq. (\ref{eq:time_bi_encoder}), \textit{i.e.}, the cross-encoder has a much higher time complexity than the bi-encoder.

\subsection{Reconstruct Cross-encoding by Bi-encoder}

To tradeoff the pros and cons of bi-encoder (Eq. (\ref{eq:bi_encoder})) and cross-encoder (Eq. (\ref{eq:cross_encoder})), here we propose a new bi-encoder architecture:
\begin{align}
\textcolor{blue}{e_D = \mathcal{M}(D)}, e_Q &= \mathcal{M}(Q), \textcolor{cyan}{e_{QD} = \mathcal{P}(e_Q, e_D)}, \textcolor{cyan}{\hat{s} = \mathcal{H}(e_{QD})} \label{eq:faratriever}
\end{align}
Similar to the nominal bi-encoder, we separately encoded $e_Q$ and $e_D$ to obtain a high computation efficiency. To obtain a comparable retrieval accuracy to the cross-encoder, however, we insert an extra latent predictor $\mathcal{P}$ to project $e_{QD}$ into a new latent space, such that $e_{QD} \rightarrow e^{\text{ce}}_{QD}$, \textit{i.e.}, to reconstruct the cross-encoded $e^{\text{ce}}_{QD}$ from Eq. (\ref{eq:cross_encoder}). Note $\mathcal{M}(D)$ can be computed offline; as long as we design $\mathcal{P}$ and $\mathcal{H}$ to be \textbf{lightweight enough}, we obtain a lower time complexity than the cross-encoder. In the next section, we will deduce an energy-based objective which matches the JEPA formulation (Eq. (\ref{eq:jepa_loss})).






\section{Method}
\label{sec:method}

In this section, we first develop a generative LLM-based retriever, which behaves as the \textbf{teacher} with quadratic time complexity. After that, we provide the methodology of the  \textbf{student}, which has a bi-encoder architecture with linear online time complexity. 


\subsection{Teacher: A Generative LLM-based Reranker}


For the \textbf{teacher}, we implement the reranker with an LLM backbone, which accepts a textual input and produces next-token probabilities:
\begin{align}
    \text{P}^{\text{teacher}} \sim \mathcal{H}^{\text{ce}} (\mathcal{M}^{\text{ce}}(Q, D)) &:= \text{LLM}(\mathcal{I}(D, Q)) \label{eq:lahore}
\end{align}
where $\mathcal{M}^{\text{ce}}$ are the LLM blocks and $\mathcal{H}^{\text{ce}}$ are the LLM language heads. According to the LaHoRe framework \cite{10.1007/978-3-031-88714-7_27}, we employ an instruction template $\mathcal{I}$ which puts the document before the query: 

\begin{tabular}{l}
\hline
\textbf{Document}: $\{D\}$ \\
\textbf{Query}: $\{Q\}$  \\
Can \textbf{Query} be appropriately replied with \textbf{Document}? \\
If the answer is true, choose \textless T\textgreater; otherwise, choose \textless F\textgreater.\\
\hline
\end{tabular}

The format of the binary-choice question follows the relevance generation paradigm \cite{sun-etal-2023-chatgpt}, which encourages LLM to thoroughly consider its preference on the relevance, and outputs either \textless T\textgreater or \textless F\textgreater, with corresponding log probabilities. Based on the causal attention mechanism of LLM, \textbf{the $D$ part of encoding is equivalent to the separately encoded $D$:}
\begin{equation}
    \text{LLM}(D, Q) [:L_D] =  \text{LLM}(D) \label{eq:casual_mechanism}
\end{equation}
where $[ ]$ indicates the indexed selection on the sequence. This property precludes the need for the $D$-encoder distillation, with only the reconstruction of $Q$ needed.

\paragraph{Training.} We construct the QA-format training samples from the original retrieval dataset, by augmenting with all possible negative pairs. The question is constructed by incorporating the query and document into the instruction template $\mathcal{I}$; while the answer is annotated by either \textless T \textgreater (for positive pair) or \textless F \textgreater (for negative pair). Then we conduct a standard supervised fine-tuning (SFT) on Eq. (\ref{eq:lahore}), with the cross-entropy loss ($\mathcal{L}$) solely on the answer. The retrieval task is then converted to a pure generative task, and the fine-tuned LLM becomes the \textbf{teacher} for the subsequent distillation. Figure \ref{fig:framework} (I) shows this training mechanism.




\paragraph{Inference.} During the inference stage, the relevance score is deduced from the \textbf{teacher}'s relative positive-negative confidence:
\begin{align}
    \hat{s}^{\text{teacher}} = \log\text{P}(\textless T \textgreater) / ( \log\text{P}(\textless T \textgreater) + \log\text{P}(\textless F \textgreater) ) \label{eq:teacher_inference} 
\end{align}



\paragraph{Time complexity.} This retriever adheres to the online time complexity of the typical \textbf{cross-encoder}: $n \text{O}(L_D + L_Q)^2$, as indicated in Eq. (\ref{eq:time_cross_encoder}). This \textcolor{red}{high time complexity} prevents the application of \textbf{teacher} as an efficient retriever, making the distillation necessary.


\begin{table*}[htbp!]
\centering
\caption{Performances on three rationale-based retrieval tasks. `R' denotes Recall. Among the rerankers, LaHoRe performs better than RankLLaMA; {\ModelName} is then distilled from LaHoRe and achieves the best accuracy over different retrievers. \textbf{Bold} indicates the best result, while \underline{underline} indicates the second best. Numbers are in percentages.}
\label{tab:result}
\small
\begin{tabular}{ll|c|ccc|ccc|cccc|c}
\toprule
 & \multicolumn{2}{c|}{Dataset($\rightarrow$)} & \multicolumn{3}{c|}{ESConv} & \multicolumn{3}{c|}{PsyQA} & \multicolumn{4}{c}{Saycan} & RT $\downarrow$ \\
\cline{2-3} \cline{4-6} \cline{7-9} \cline{10-13} \cline{14-14}
 & Method ($\downarrow$) & Size & R@1 & R@3 & MRR & R@1 & R@3 & MRR & R@1 & R@3 & R@5 & MRR & (ms) \\
\midrule
\parbox[t]{5mm}{\multirow{10}{*}{\rotatebox[origin=c]{90}{
\parbox[c]{2cm}{\centering \scriptsize Retriever}}}} 
& TART-Contriever \cite{asai2023tart} & 110M & 1.4 & 38.5 & 35.1 & 11.3 & 44.2 & 36.5 & 1.2 & 19.2 & 36.5 & 17.5 \\
& E5-large-instruct \cite{wang2024E5} & 550M & 19.2 & 47.2 & 40.5 & 12.9 & 42.8 & 36.3 & 7.7 & 17.3 & 26.9 & 18.2 \\
& InstructOR XL \cite{su2023instructorXL} & 1.5B & 13.7 & 44.3 & 36.4 & 24.6 & 61.2 & 48.1 & 17.3 & 26.9 & 34.6 & 28.4 \\
& RepLLaMA \cite{ma2024RepLLaMA} & 7B & 17.4 & 49.1 & 39.7 & 29.6 & 56.0 & 49.6 & 9.6 & 25.0 & 38.4 & 23.6 \\  
& Promptriever \cite{weller2024promptrieverinstructiontrainedretrieversprompted} & 7B & 17.6 & 48.9 & 39.5 & 29.6 & 61.7 & 50.9 & 5.8 & 19.2 & 34.6 & 19.0 \\ 
& E5-Mistral \cite{wang2024E5} & 7B & 15.5 & 50.2 & 39.3 & 11.4 & 48.0 & 36.8 & 14.9 & 19.1 & 19.1 & 17.0 \\
& GritLM \cite{muennighoff2024GritLM} & 7B & 19.2 & 50.2 & 41.7 & 26.9 & 57.5 & 48.4 & 35.8 & 43.5 & 43.1 & 46.3 \\ 
& OneGen \cite{zhang2024onegen} & 7B & \underline{29.0}  & 52.6 & 41.5 & 31.3 & 46.0 & 46.9 & \underline{44.2} & \underline{67.3} & \underline{75.0} & \underline{56.8} &  \\
& LlaMA2Vec \cite{li2024llama2vec} & 7B & 23.9 & \underline{55.6} & \underline{45.5} & \bf 34.3 & \underline{75.4} & \bf 57.5 & 17.3 & 40.4 & 46.2 & 31.1 & \bf 93 \\ 
& \textbf{\ModelName} (ours) & 7B & \bf 32.8 & \bf 66.4 & \bf 54.7 & \underline{32.8} & \bf 76.1 & \underline{55.1} & \textbf{67.3} & \textbf{76.9} & \textbf{78.8} & \textbf{73.7} & \underline{95} \\ 
\midrule
\parbox[t]{5mm}{\multirow{2}{*}{\rotatebox[origin=c]{90}{
\parbox[c]{0.8cm}{\centering \scriptsize Reranker}}}}
& RankLLaMA \cite{ma2024RepLLaMA} & 7B & 32.3 & 64.8 & 52.9 & 33.6 & 77.4 & 57.2 & 26.9 & 50.0 & 75.0 & 44.9 \\
\cmidrule{2-13}
& LaHoRe \cite{10.1007/978-3-031-88714-7_27} (\textbf{teacher}) & 7B & \bf 34.4 & \bf 67.1 & \bf 54.8 & \bf 35.6 & \bf 77.6 & \bf 58.2 & \bf 71.2 & \bf 88.5 & \bf 94.2 & \bf 79.8 & 980 \\
\bottomrule
\end{tabular}
\end{table*}

\subsection{On-policy Distillation by JEPA}

\paragraph{The Architecture.} As a bi-encoder \textbf{student}, we implement both $D$ and $Q$ encoders with the same LLM backbone, with parameters initialized from the \textbf{teacher}, \textit{i.e.}, \textit{i.e.}, $\mathcal{M} \leftarrow \mathcal{M}^{\text{ce}}$, $\mathcal{H} \leftarrow \mathcal{H}^{\text{ce}}$ according to Eq. (\ref{eq:faratriever}) and Eq. (\ref{eq:cross_encoder}). Given $\textcolor{blue}{e_D = \mathcal{M}}(D), e_Q = \mathcal{M}(Q)$\footnote{Note $e_Q$ is equivalent with $\text{LLM}(D)$ in Eq. (\ref{eq:casual_mechanism}) due to the causal mechanism of LLM.}, we then insert a lightweight predictor block $\mathcal{P}$ between $\mathcal{M}$ and $\mathcal{H}$, which produces a new embedding and generates the final retrieval score:
\begin{align}
    &e^{\text{student}}_Q = \textcolor{cyan}{\mathcal{P}_{z=e_D}(e_Q)}, \quad \text{P}^{\text{student}} \sim \textcolor{cyan}{\mathcal{H}(e_D \oplus e^{\text{student}}_Q)} \label{eq:predictor}
\end{align}
where $z$ is the latent variable of $\mathcal{P}$ to impose the grounding impact of $D$ on $e_Q$, and $\oplus$ means vector concatenation. The relevance score $\hat{s}^{\text{student}}$ is calculated from $\text{P}^{\text{student}}$ the same as Eq. (\ref{eq:teacher_inference}). 

 
\paragraph{Connection to JEPA.} By minimizing the energy between \textbf{teacher} and \textbf{student} embedding, the architecture can be considered as the JEPA paradigm:
\begin{align}
    \min_{\mathcal{P}}{\mathcal{F}(e_D \oplus \mathcal{P}_{z=e_D}(e_Q), e^{\text{ce}}_{QD})} \label{eq:objective} 
\end{align}
Compared to the original formulation of JEPA, \textit{i.e.}, Eq. (\ref{eq:jepa_loss}), $e_Q$ plays as the input $x$, $e^{\text{ce}}_{QD}$ plays as the output $y$, and $e_D$ is the latent $z$. Figure \ref{fig:paradigm} visualizes this distillation paradigm.


\paragraph{The Predictor.} We implement $\mathcal{P}$ first with a 2-layer MLP with Relu activation, then applying a light-weight elementary multiplication:
\begin{equation}
    e^{\text{student}}_Q = \text{MLP}_{\theta} (e_Q) \odot e_D[-1] 
\end{equation}
where $[-1]$ denotes the last embedding of the sequence. During \textbf{student} training, only the MLP parameter $\theta$ is trainable. Although all the modules of $\mathcal{P}$ seems simple, later experiments show that they are effective enough for the distillation. Apparently, they are lightweight enough to ensure inference efficiency.

\paragraph{Training.} We replace the next-token prediction in the \textbf{teacher} with the energy-based modeling (EBM), by minimizing the energy between \textbf{teacher} and \textbf{student} on the latent space. As suggested by JEPA \cite{841b6929fdd44bd58434b6ac370439e2}, we implement mean-square error (MSE) instead of a contrastive loss, which helps improve the convergence and stability of EBM. Besides MSE on $Q$ embedding, we augment an auxiliary reverse KL of \textbf{student} w.r.t \textbf{teacher}, which further pushes the \textbf{student} to approximate the \textbf{teacher}'s behavior on the logit level:
\begin{align}
    \min_{\theta}{\mathcal{F}} := &\mathcal{F}^{\text{MSE}}(e^{\text{student}}_Q, e^{\text{teacher}}_Q) + w KL^{\text{rev}}(\text{P}^{\text{student}} || \text{P}^{\text{teacher}}) \label{eq:student_objective} 
\end{align}
where $w$ is the KL loss weight.  Apparently, the energy $\mathcal{F}$ decays to zero when \textbf{student} perfectly reconstructs \textbf{teacher}. In the ablation study, we verify that both $\mathcal{F}^{\text{MSE}}$ and $KL^{\text{rev}}$ are critical for retrieval performance improvement. 

During the training, the student is actively updating, rollouts all samples while the teacher is frozen and infers $e^{\text{teacher}}_Q$ and $\text{P}^{\text{teacher}}$, which corresponds to the on-policy distillation. The architecture, loss, and training mechanism are visualized in Figure \ref{fig:framework} (III).





\subsection{Inference Efficiency}
\label{sec:inference_efficiency}

Within the predictor, the time complexity of MLP is $\text{O}(L_Q)$ while the dot multiplication is $\text{O}(1)$. The entire time complexity of {\ModelName} then becomes
\begin{align}
    \underbrace{\textcolor{blue}{n \text{O}(L_D)^2}}_{\text{Offline}} + \underbrace{\text{O}(L_Q)^2 + \textcolor{cyan}{\text{O}(L_Q) + n \text{O}(1)}}_{\text{Online}} \label{eq:time_student_inference}
\end{align} 
Although slower than the original bi-encoder method (Eq. (\ref{eq:time_bi_encoder}) has only linear components), apparently, its online time complexity has been much improved from pure cross-encoder methods (Eq. (\ref{eq:time_cross_encoder})).


\section{Experiment}
\label{sec:experiment}

In this section, we first introduce the experimental settings, including baselines, metrics, and implementation details. Then we conduct experiments on rational-based retrieval tasks, followed by detailed ablation and efficiency studies. Finally, we investigate {\ModelName}'s performance on traditional retrieval benchmarks.

\subsection{Settings}

\paragraph{Implementation.} We set the maximum of $L_D$ to 32 and conduct \textbf{left-padding} to the document part. The rest of the textual input is \textbf{right-padded}, with the total window length of $4096$. {\ModelName} is initialized from Qwen2-7B-Instruct \cite{qwen2techreport2023}. Training is conducted using LlamaFactory \cite{zheng2024llamafactory}, with 8 A100 GPUs and takes less than 10 hours. The learning rate is $1.0e-6$, the batch size is 16, and the epoch number of epochs The KL loss weight is set to $w=0.25$. Inference is implemented with vLLM with the prefix cache \cite{kwon2023vLLM}. For the efficiency analysis, the online response time (RT) is calculated and recorded for each query-document pair in the test set, and the average is reported. Inspired by \cite{10.1007/978-3-031-88714-7_27}, for both \textbf{teacher} and \textbf{student}, we construct the dataset by augmenting the positive query-document pairs with all possible negative pairs to leverage the scaling of LLM. 


\paragraph{Metrics.} We evaluate on different retrieval metrics, including Recall@k (k may be set to 1, 3, 5, considering the total number of documents), p-MRR@10, as well as nDCG@10 for ranking-annotated benchmarks such as BEIR. In the following contexts, we use R@k to denote Recall@k, while using MRR to abbreviate p-MRR@10.


\paragraph{Baselines.}
We compare {\ModelName} to typical retrievers based on different architectures or methodologies, including traditional dense retrievers such as \textit{TART-Contriever} \cite{asai2023tart}, \textit{RepLLaMA} \cite{ma2024RepLLaMA}, \textit{E5-large-instruct}, and \textit{E5-Mistral} \cite{wang2024E5}; instructional retrievers such as \textit{InstructOR XL} \cite{su2023instructorXL}, \textit{Promptriever} \cite{weller2024promptrieverinstructiontrainedretrieversprompted} and \textit{LLaMA2Vec} \cite{li2024llama2vec}; and multi-task retrievers like \textit{GritLM} \cite{muennighoff2024GritLM} and \textit{OneGen} \cite{zhang2024onegen}, which integrates the representative and generative tasks. For rerankers, we experiment with \textit{RankLLaMA}, the cross-encoder variant of \textit{RepLLaMA} \cite{ma2024RepLLaMA}; as well as \textit{LaHoRe} \cite{10.1007/978-3-031-88714-7_27}, which behaves as the \textbf{teacher} of distillation in this paper. For ease of comparison, most baselines have an identical parameter size (7B) to our implementation, as highlighted in Table~\ref{tab:result}.

\subsection{Rationale-based Retrieval}

\paragraph{Datasets.} In this subsection, we conduct the experiments on two typical scenarios of rational-based retrieval:

\noindent \textit{(1) Empathetic conversation.} These datasets focus on emotional support and psychological counseling, where the system responses are annotated with predefined, optimal support strategies.
\begin{itemize}
    \item \textbf{ESConv} \cite{liu2021ESconv} is a multi-turn emotional support dialogue dataset containing 1.2k conversations and 12.7k utterances. Each supporter response is annotated with one of eight fine-grained support strategies, enabling the study of long-horizon strategy planning in empathetic conversations.
    \item \textbf{PsyQA} \cite{sun-etal-2021-psyqa} is a single-turn psychological counseling dataset consisting of 4,012 question--answer pairs. Each response is labeled with one of seven counseling strategies, emphasizing the strategy selection under limited conversational contexts.
\end{itemize}
For both datasets, the query is formalized by prepending the historical context (if any) to each user utterance, while all strategy candidates form the pool of documents. 

\noindent \textit{(2) Robotic manipulation.} In this scenario, the robot makes decisions via language grounding, each time choosing the best subskill from all textual skill candidates, in order to achieve its long-term goal.

\begin{itemize}
    \item \textbf{SayCan} \cite{pmlr-v205-ichter23a} is a large-scale robotic manipulation dataset with approximately 276K trajectories, each consisting of up to 50 steps and covering 551 low-level skills. The dataset supports research on grounding natural language instructions into executable action sequences through hierarchical planning.
\end{itemize}

\paragraph{Retrieval Results.} Table \ref{tab:result} shows retrieval performances on three rationale-based datasets. Although fast, most of retriever baselines fail to capture cross-semantic relations between queries and documents, with generally lower accuracies. Although this issue can be alleviated by integrating a generative subtask, such as GritLM and OneGen; or prompting the LLM for next-sentence inference, as did by LlaMA2Vec, they still perform worse than LaHoRe, the cross-encoding \textbf{teacher}.  In contrast, our \textbf{\ModelName} improves the performance significantly, revealing the effectiveness of distillation.









\begin{table}[h!]
\centering
\caption{Ablation studies on PsyQA and SayCan.} 
\label{tab:ablation_result}
\small
\begin{tabular}{l|ccc|cccc}
\toprule
 \multicolumn{1}{c|}{Dataset($\rightarrow$)} & \multicolumn{3}{c|}{PsyQA} & \multicolumn{4}{c}{SayCan} \\ 
\cline{1-1} \cline{2-4} \cline{5-8}
 Method ($\downarrow$) & R@1 & R@3 & MRR & R@1 & R@3 & R@5 & MRR \\ 
\midrule
\textit{w/o-}$mul$ & 2.0 & 10.7 & 21.3 & 9.6 & 15.4 & 19.2  & 14.7 \\
\textit{w/o-}$\mathcal{F}^{\text{MSE}}$ & 0.0 & 0.7 & 17.2 & 23.1 & 40.4 & 44.2 & 33.1 \\
\textit{w/o-}$KL^{\text{rev}}$ & 24.1 & 24.9 & 37.3 & 21.2 & 30.8 & 46.2 & 31.8 \\
\textit{w/-}$KL^{\text{forward}}$ & 28.4 & 55.3 & 48.5 & 46.8 & 54.1 & 62.4 & 59.7 \\
\bf {\ModelName} & \textbf{32.8} & \textbf{76.1} & \textbf{55.1} & \textbf{67.3} & \textbf{76.9} & \textbf{78.8} & \textbf{73.7} \\ 
\bottomrule 
\end{tabular}
\end{table}

\paragraph{Compared to the teacher and bi-encoders.} An important research question can be further proposed: \textbf{if {\ModelName} effectively trades off the accuracy of reranker and the computational efficiency of bi-encoder retrievers?} From Table \ref{tab:result}, one can observe that {\ModelName}'s performance is relatively close to \textbf{teacher}, indicating the knowledge is effectively preserved during the distillation. For Saycan, {\ModelName} even apparently outperforms another reranker, RankLLaMA. Table \ref{tab:result} also compares the average response time (RT) of {\ModelName}, \textbf{teacher}, and the best retriever baseline, LlaMA2Vec, in milliseconds. {\ModelName} has a similar RT to LlaMA2Vec, while is much faster than \textbf{teacher}, which validates its retrieval efficiency.


\paragraph{Ablation Study.} To verify the necessity of method components, we first attempt to remove the MSE loss (\textit{w/o-$\mathcal{F}^{\text{MSE}}$}), the KL divergence (\textit{w/o-$KL^{\text{rev}}$}), and the multiplication operator (\textit{w/o-mul}). We also replace the reverse KL with forward KL (\textit{w/-$KL^{\text{forward}}$}), to validate the benefit of reverse KL within the on-policy distillation.

Table \ref{tab:ablation_result} indicates that {\ModelName} outperforms all these ablations, validating the reasonability of our framework. Especially, \textit{w/o-mul} completely fails, because the multiplication operation introduces the grounding information of $D$ on $Q$, which is critical to reconstruct the cross-encoding distribution. Removing the reverse KL term, or switching it to the forward KL, resulting in significant performance degradation, verifying that reverse KL significantly helps the \textbf{student} focus on the \textbf{teacher}'s critical behaviors (answer True or False solely), facilitating the convergence by conditioning on \textbf{student}'s own distribution. Finally, removing the latent difference loss ($\mathcal{F}^{\text{MSE}}$) resulting in varied impacts, where the PsyQA performance is much more affected than SayCan. We suppose the reason is that PsyQA infers much longer context window (with the conversation history) than SayCan, in which situations distillation on the latent space becomes more critical for model convergence.

\begin{figure}[htbp!]
\centering
  \includegraphics[width=1\linewidth]{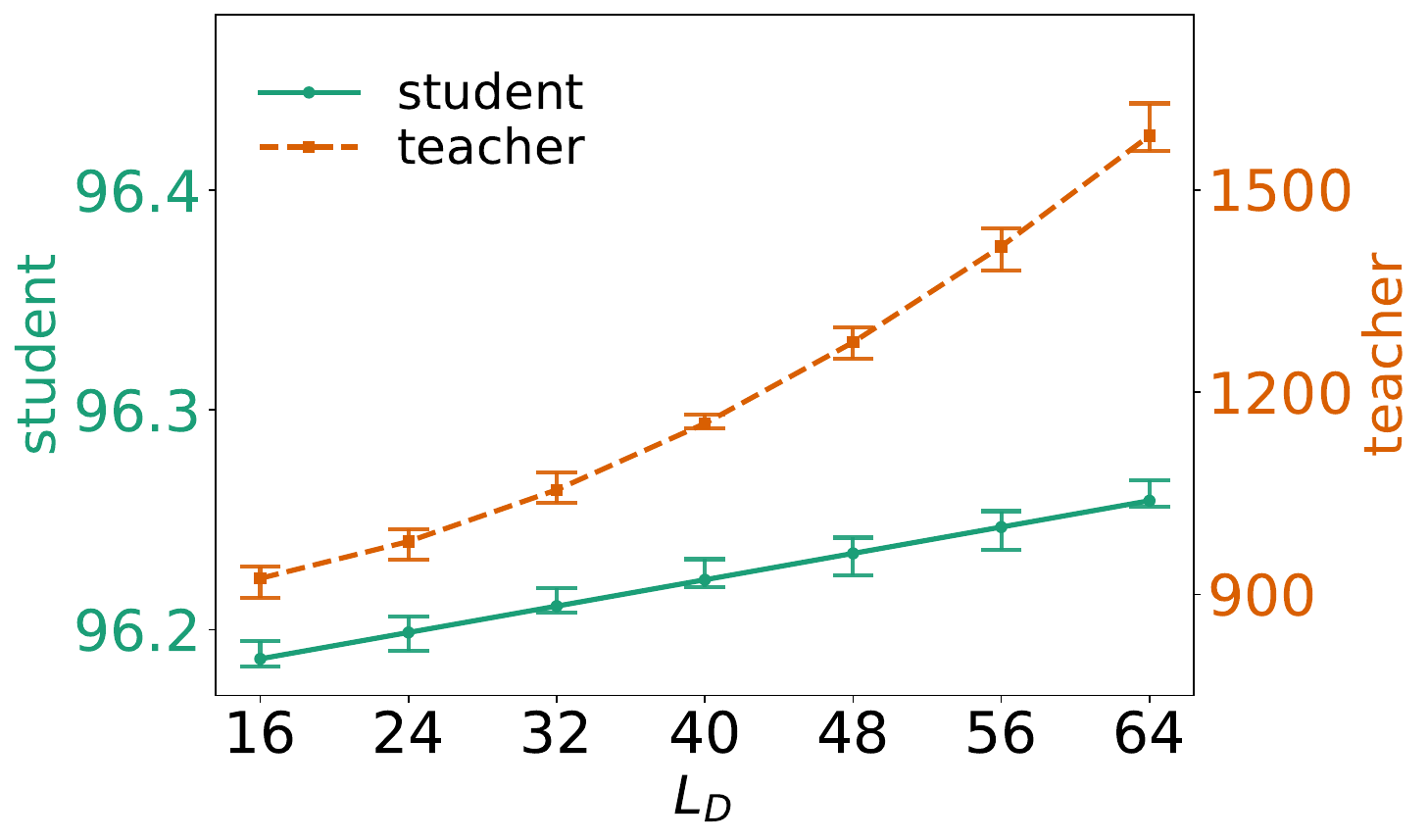}
  \caption{Averaged Online RTs corresponding to different $L_D$.} 
  \label{fig:sensitivity3}
\end{figure}

\paragraph{The RT dependency.} As derived in Eq. (\ref{eq:time_student_inference}), our theoretical efficiency analysis indicates that {\ModelName} achieves a \textit{linear} dependency on the document length ($L_D$), in contrast with the \textit{quadratic} dependency of cross-encoder \textbf{teacher} (Eq. (\ref{eq:time_cross_encoder})). To provide the empirical verifications, we provide the statistics of online RTs with respect to different values of $L_D$. In Figure \ref{fig:sensitivity3}, one can clearly observe the \text{linear} dependency of \textbf{teacher} and the quadratic dependency of \textbf{teacher} on $L_D$, verifying the theoretical conclusion. 

\begin{figure}[htbp!]
\centering
  \includegraphics[width=0.85\linewidth]{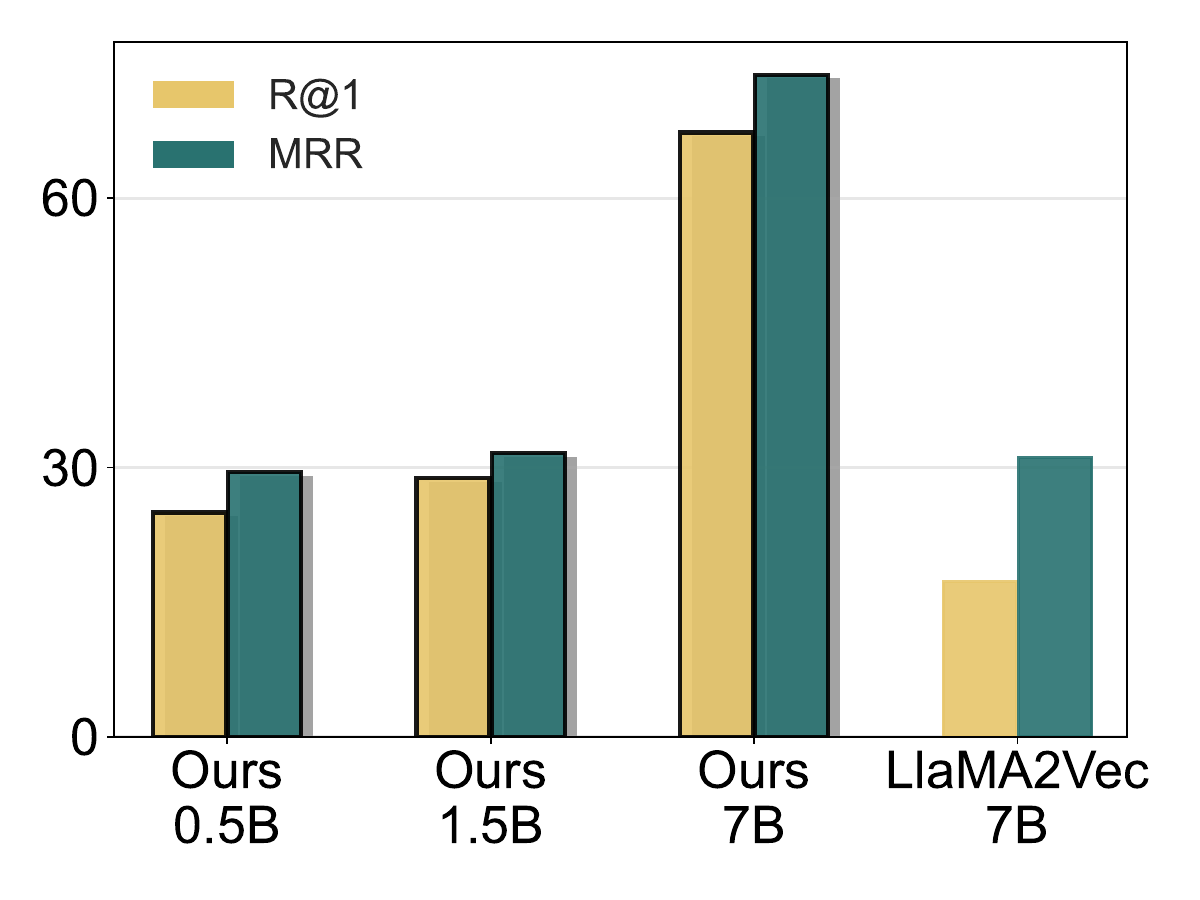}
  \caption{Method scalability on SayCan.} 
  \label{fig:scalability}
\end{figure}

\paragraph{Scalability.} Figure \ref{fig:scalability} exhibits the SayCan performance of {\ModelName} on smaller backbones, including 0.5B and 1.5B. Compared to LlaMA2Vec which has a 7B backbone, {\ModelName} performs similarly, even with much smaller backbones. This promising result not only verifies the scalability of {\ModelName}, but also reveals the possibility of further improving the retrieval speed at a tradeoff with performance, by using smaller backbones if necessary.



\subsection{Generalize to Fact-based Retrieval}

To verify the generalization capability of {\ModelName}, we further test it on two famous fact-based retrieval benchmarks, comparing with current SoTA baselines.

\paragraph{Settings.}  For in-domain test, we evaluate on MS MARCO \citep{nguyen2016ms} passage retrieval tasks, which consists of approximately 500k training examples. MRR@10 are reported on the \textit{dev} split. For out-of-domain test, models trained by MS MARCO are evaluated on BEIR \citep{thakur2021beir}, in a zero-shot manner. We test on the 13 subtasks of BEIR which are publicly available, and report the nDCG@10 results.

\begin{figure*}[htbp!]
\centering
  \includegraphics[width=0.8\linewidth]{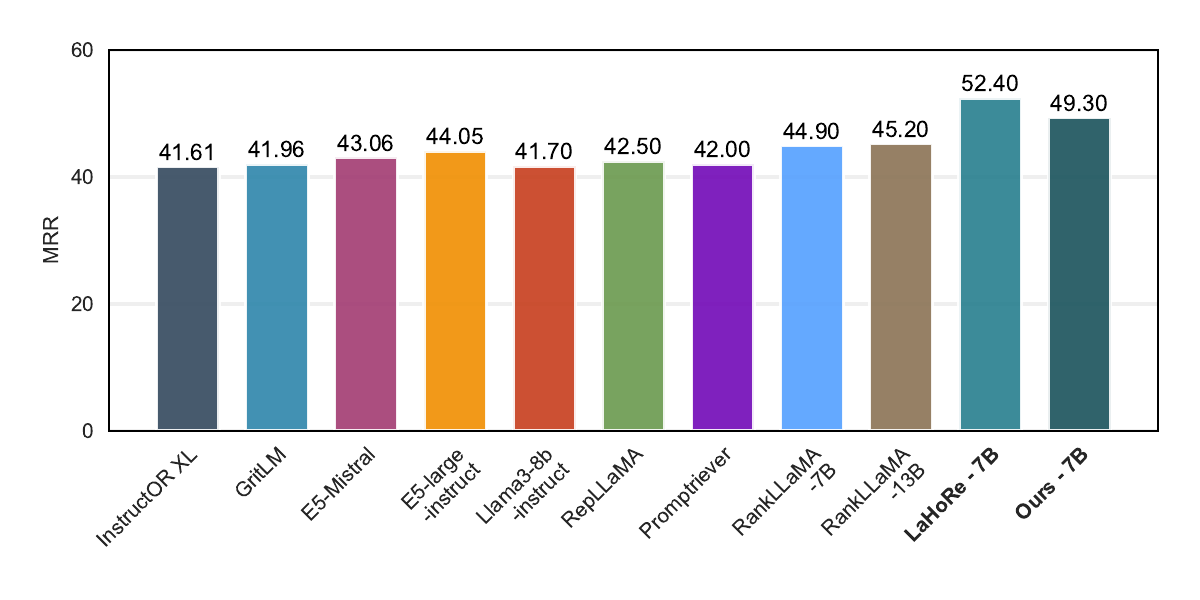}
  \vspace{-0.1in}
  \caption{MRR comparison of models on MS MARCO passage retrieval.}
  \label{fig:msmarco}
\end{figure*}

\paragraph{In-domain performance.} Figure \ref{fig:msmarco} compares the MS MARCO results of different methods. The \textbf{teacher} outperforms all baselines with MRR of 52.4, indicating stronger comprehension from cross-encoding rerankers. Furthermore, our \textbf{student} also obtains an MRR of 49.30, surpassing all retriever baselines (although less than \textbf{teacher}). This promising result shows that our distillation methodology can also effectively adapt to generalized retrieval tasks, while improving the inference speed of the cross-encoder dramatically.

\begin{table*}[t!]
\caption{Zero-shot performance on BEIR (nDCG@10). Best value in the row is bolded. The best result is marked \textbf{bolded} and the second best is \underline{underlined}. Compared to the best reranker LaHoRe, {\ModelName} obtains averagely 58.2, and performs the best in 2/13 subtasks and second-best in 4/13 subtasks, which is close to the best retriever competitor (GritLM). }
\label{tab:beir}

\setlength\tabcolsep{3 pt} %
\centering
\small
\begin{tabular}{ll|ccccccc|cc|c}
\toprule
& \multirow{3}{*}{Dataset} & \multicolumn{7}{c|}{Retriever} & \multicolumn{2}{c|}{Reranker} & \multicolumn{1}{c}{\textbf{Ours}} \\
\cmidrule(l){3-9} \cmidrule(l){10-11} \cmidrule(l){12-12}
 & & \multirow{2}{*}{\shortstack{BM25\\N/A}} & \multirow{2}{*}{\shortstack{E5-large\\550M}} & \multirow{2}{*}{\shortstack{E5-Mistral\\7B}} & \multirow{2}{*}{\shortstack{Promptriever\\7B}} & \multirow{2}{*}{\shortstack{RepLLaMA\\7B}} & \multirow{2}{*}{\shortstack{LlaMA2Vec\\7B}} & \multirow{2}{*}{\shortstack{GritLM\\7B}} & \multirow{2}{*}{\shortstack{RankLLaMA\\7B}} & \multirow{2}{*}{\shortstack{LaHoRe\\7B}} & \multirow{2}{*}{\shortstack{\ModelName\\7B}} \\
 \\ 
\midrule
\parbox[t]{3mm}{\multirow{7}{*}{\rotatebox[origin=c]{90}{
\parbox[c]{2.5cm}{\centering \scriptsize Dataset has dev/train set}}}}
& DBPedia & 29.9  & 41.3 & 48.9 & 45.2 & 44.8 & 45.9 & 46.6 & 48.3 & \bf 75.5 & \underline{62.7} \\
& FEVER & 48.1 & 82.8 & \bf 87.8 & 82.8 & 82.9 & 81.3 & 82.7 & 83.9 & \underline{84.9} & 76.7 \\
& FiQA & 25.1 & 43.8 & 56.6 & 46.6 & 45.0 & 48.5 & \underline{60.0} & 46.5 & \bf 91.4 & 54.9 \\
& HotpotQA & 56.9 & 71.2 & 75.7 & 69.5 & 68.8 & 70.1 & \underline{79.4} & 75.3 & 76.9 & \bf 80.0 \\
& NFCorpus & 32.1 & 34.0 & 38.6 & 36.9 & 36.0 & 38.2 & \bf 41.2 & 30.3 & \underline{40.6} & \underline{40.6} \\
& Quora & 80.4 & 88.2 & \bf 89.6 & 88.0 & 86.0 & 88.3 & \underline{89.5} & 85.0 & 88.2 & 88.5 \\
& SciFact & 68.7 & 70.4 & 76.4 & 76.3 & 75.3 & 74.8 & \bf 79.2 & 73.2 & \underline{78.9} & \underline{78.9}\\
\midrule
\parbox[t]{5mm}{\multirow{6}{*}{\rotatebox[origin=c]{90}{
\parbox[c]{2cm}{\centering \scriptsize No dev/train set}}}}  
& Arguana & 36.6 & 54.4 & \underline{61.9} & 56.7 & 48.6 & 56.5 & \bf 63.2 & 56.0 & 36.8 & 36.8 \\
& Climate-FEVER & 13.6 & 25.7 & \underline{38.4} & 32.1 & 29.3 & 38.2 & 30.9 & 28.0 & \bf 42.3 & 23.7 \\
& NQ & 28.5 & 64.1 & 63.5 & 62.6 & 63.0 & 64.6 & 70.3 & 66.3 & \bf 92.0 & \underline{73.1} \\
& SCIDOCS & 15.8 & 17.5 & 16.3 & 19.7 & 16.1 & 18.9 & \underline{24.4} & 17.8 & 17.1 & \bf 25.0 \\
& TREC-COVID & 62.3 & 71.3 & 87.2 & 84.6 & 83.9 & \bf 86.9 & 74.8 & \underline{85.2} & 81.6 & 81.8 \\
& Touche-2020 & 33.1 & 23.4 & 26.4 & 32.0 & 34.1 & 34.2 & 27.9 & \underline{40.1} & \bf 49.9 & 34.3 \\
\midrule
& \# of best & 0 & 0 & 2 & 0 & 0 & 1 & 3 & 0 & 5 & 2 \\
& \# of 2nd-best & 0 & 0 & 2 & 0 & 0 & 0 & 4 & 2 & 3 & 4 \\
& Average & 40.9 & 53.9 & 59.0 & 56.4 & 54.9 & 57.4 & \underline{59.2} & 56.6 & \bf 65.8 & 58.2 \\
\bottomrule
\end{tabular}
\end{table*}

\paragraph{Zero-shot results.} We finally evaluate retrievers trained by MS MARCO on BEIR, with results in Table \ref{tab:beir}. Again, the \textbf{teacher} obtained the best averaged nDCG@10 across all retrievers, performing the best on 5/13 subtasks and the second best on 3/13 subtasks. On the other hand, the \textbf{student} obtains a comparable performance to the cross-encoder rerankers, and has the second-best averaged result among all bi-encoder retrievers. Especially, the \textbf{student} performs the best on 2/13 subtasks and the second best on 4/13 subtasks, which is close to GritLM, the most competitive bi-encoder baseline.


\section{Conclusion}
\label{sec:conclusion}


In this paper, we propose an efficient, LLM-based retriever called {\ModelName}, to address the challenging problem of rationale-based retrieval. We balance the retrieval accuracy and the computational efficiency by implementing a bi-encoder architecture, while providing a retrieval score that considers the cross-semantic information between query and document. We achieve this target by on-policy distilling from a cross-encoder reranker, based on the JEPA framework, with both latent MSE and reverse KL losses implemented. Experiments verify that our {\ModelName} can handle rationale-based retrieval problems, surpassing both bi-encoder and cross-encoder baselines, and achieve a much faster speed than a pure LLM-based cross-encoding retriever. We further generalize {\ModelName} to traditional fact-based retrieval tasks, and also observe state-of-the-art performances obtained on both MS MARCO and BEIR benchmarks.


\clearpage
\newpage

\bibliographystyle{ACM-Reference-Format}
\bibliography{main}

\clearpage
\newpage

\appendix

\section{More Implementation Details}

\subsection{Detailed Prompt}

Here we present the detailed instruction prompt $\mathcal{I}$:

\begin{tcolorbox}[
    title=The instruction prompt $\mathcal{I}$, 
    colback=white, 
    colframe=pink!75!black, 
    colbacktitle=pink, 
    coltitle=black, 
    label=tc:prompt, 
    fonttitle=\bfseries  
]
\textbf{Document}: $\{D\}$ \\
\textbf{Query}: $\{Q\}$  \\
Can \textbf{Query} be appropriately responded with \textbf{Document}? \\
If the answer is true, choose \textless T\textgreater; otherwise, choose \textless F\textgreater.
\end{tcolorbox}
The first part connects $D$ and $Q$ from the sense of semantic relation, and the second part encourages LLM to thoroughly consider its preference on the relevance by augmenting a binary-choice problem, and finally provides the answer. The instruction encourages LLM to thoroughly consider its preference on the relevance, and outputs either \textless T \textgreater or \textless F \textgreater, with corresponding log probabilities.

\subsection{Comparison of Retriever Paradigms}

Table \ref{tab:overview} summarizes three paradigms. In the table (and also the following formulations), we use \textcolor{blue}{blue} to denote modules which can be calculated off-the-shelf (offline); and use \textcolor{cyan}{cyan} to denote lightweight calculations.

We then formulate different encoders as follows:
\begin{itemize}
    \item the classical dual-encoder:
    \begin{equation}
        s \sim \textcolor{cyan}{\mathcal{H}(s|e_Q, e_D)} \mathcal{M}^Q(e_Q | Q) \textcolor{blue}{\mathcal{M}^D(e_D | D)} \label{eq:dual-encoder}
    \end{equation}
    \item the cross-encoder:
    \begin{equation}
        s^{\text{ce}} \sim \textcolor{cyan}{\mathcal{H}^{\text{ce}}(s|e_{QD})} \mathcal{M}^{\text{ce}}(e_{QD} | Q, D) \label{eq:cross-encoder}
    \end{equation}
    \item our {\ModelName}:
    \begin{align}
         &\hat{s} \sim \textcolor{cyan}{\mathcal{H}^{\text{ce}}(\hat{s}|\hat{e}_{QD})} \textcolor{cyan}{\mathcal{P}(\hat{e}_{QD} | e_D, e_Q)} \notag \\
         &\quad \mathcal{M}(e_Q | Q) \textcolor{blue}{\mathcal{M}(e_D | D)} \label{eq:faratriever} \\
    &w.r.t \notag \\
    &\min_{\mathcal{P}} \mathcal{F}(\hat{s}, s^{\text{ce}}), \mathcal{F} \rightarrow 0, \text{ when } \hat{s} \rightarrow s^{\text{ce}} \label{eq:objective}
    \end{align}
\end{itemize}
in which $e$ is embedding vector, $\mathcal{M}$ is the encoder, $\mathcal{H}$ is the score head; the subscript $\text{ce}$ denotes the cross-encoder.

\subsection{Data Example}

One can refer to Table \ref{tab:ex_esconv} for an example from \textit{ESConv}. \textit{PsyQA} has a similar format.

\begin{table}[htbp!]
    \centering
    \small
    \begin{tabular}{c|l}
        \toprule
        All Strategies & \makecell[l]{[Question,Restatement or Paraphrasing,Self-disclosure,\\Affirmation and Reassurance,Providing Suggestions,\\Reflection of feelings,Information, Others
        ]}  \\
        \toprule
        Query & \makecell[l]{\textit{\{history\}} \\
        \textit{user:} Seriously! What I am scared of now is how to \\secure another job.}  \\
\midrule
       Strategy & \makecell[l]{\textcolor{blue}{Reflection of feelings}}   \\
\midrule
      Response & \makecell[l]{\textit{assistant:} I can feel your pain just by chatting with you.}   \\
        \bottomrule
    \end{tabular}
    \caption{An example of \textit{ESConv} as a retrieval dataset. `All Strategies' corresponds to the full document list, and `Strategy' is the ground truth of the retrieved document.}
    \label{tab:ex_esconv}
\end{table}

\subsection{Baseline Details}

Here we provide the detailed introductions of the baseline retrievers and rerankers:

\begin{itemize}
    \item \textbf{TART}-Contriever \cite{asai2023tart} is a task-aware dual-encoder retriever trained with natural language instructions to adapt retrieval behavior across tasks. 
    
    \item \textbf{InstructOR XL} \cite{su2023instructorXL} is an instruction-tuned dual-encoder that jointly encodes task instructions and text to produce task-specific embeddings. 
    
    \item \textbf{E5} \cite{wang2024E5} is a family of general-purpose dual-encoder embedding models trained with large-scale weakly supervised contrastive learning. Detailed implementations include E5-Mistral and E5-large-instruct. 

    \item \textbf{RepLLaMA} \cite{ma2024RepLLaMA} adapts LLaMA into a dual-encoder retriever using parameter-efficient fine-tuning methods such as LoRA. It also has a cross-encoder variant called \textbf{RankLLaMA}.

    \item \textbf{Promptriever} \cite{weller2024promptrieverinstructiontrainedretrieversprompted} adds the prompt mechanism on RepLLama to improve the performance on instructed retrieval.
    
    \item \textbf{OneGen} \cite{zhang2024onegen} proposes a dual-encoder paradigm that provides a unified framework for both representation learning and generative learning.
    
    \item \textbf{GritLM} \cite{muennighoff2024GritLM} trains a retriever with generative representational instruction tuning.
    
    \item \textbf{LLaMA2Vec} \cite{li2024llama2vec} derives dense representations of query and document by prompting either `SELF' or `NEXT'. 
\end{itemize}

\section{Extra Experiment Results}

\subsection{Comparison with Classifications}

\begin{figure}[t!]
\centering
\begin{minipage}[t]{0.55\linewidth}
  \vspace{0pt}  
  \includegraphics[width=\linewidth]{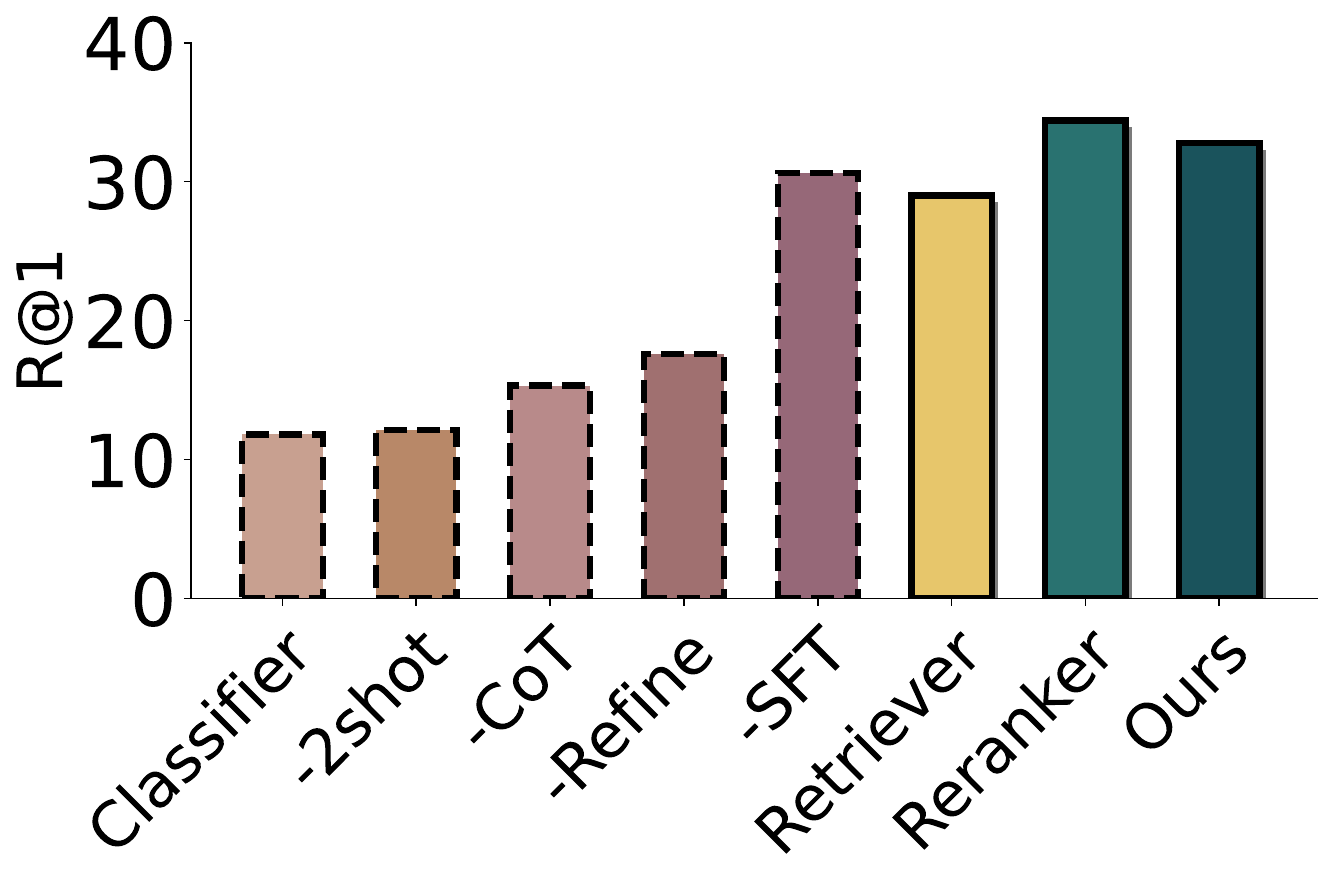}
\end{minipage}%
\hfill
\begin{minipage}[t]{0.4\linewidth}
  \vspace{0pt}  
  \includegraphics[width=\linewidth]{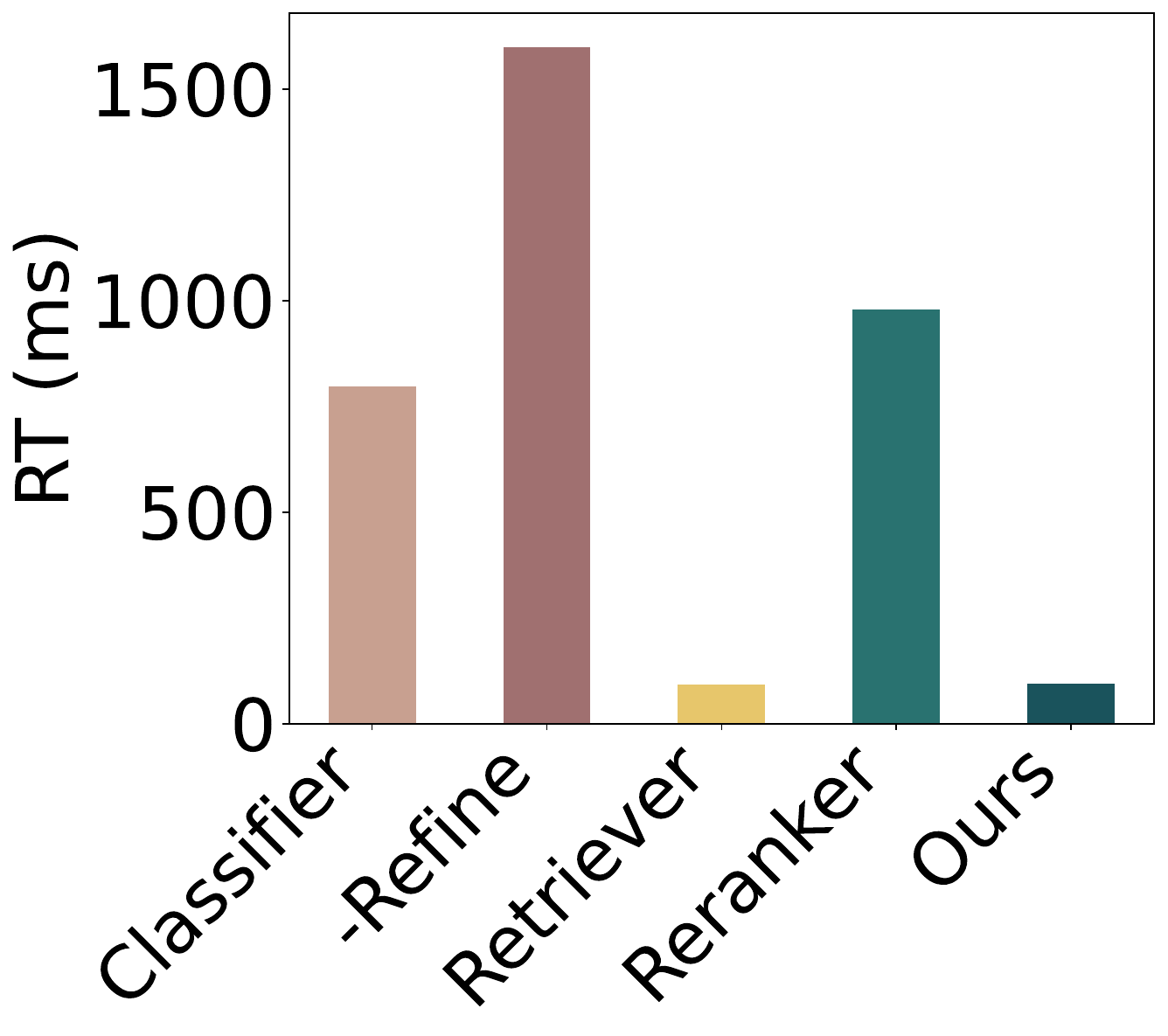}
\end{minipage}
\caption{Comparison between {\ModelName} and more textual methods. {\ModelName} outperforms the zero-shot Classifier apparently, as well as its prompting and finetuning variants. On the other hand, {\ModelName} has a similar speed to Retriever, while is much faster than Classifier-based methods.}
\label{fig:snapshot2}
\end{figure}

As a capable zero-shot learner, LLM can also be employed as a classifier for textual retrieval tasks, especially when the number of documents is relatively limited (as in the case of ESConv and PsyQA). In such a case, the researcher may ask the LLM to select the optimal document by including the contents of all the document candidates in the prompt. To verify that our {\ModelName} is also a competitive solution compared to the Classifier, we include more classifier-based baselines on SayCan, including:
\begin{itemize}
    \item Classifier: inference the LLM with the classification prompt in a zero-shot manner.
    \item Classifier-2shot: employ the in-context-learning with random two examples.
    \item Classifier-CoT: add the chain-of-thought \citep{wei2022chain} instruction into the prompt, \textit{i.e.}, `Let's think step by step'.
    \item Classifier-Self-Refine: integrate the Self-Refine method \citep{Madaan2023SelfRefine}, which infers the LLM the second time to reflect and refine the answer.
    \item Classifier-SFT: conduct the instruction tuning on the Classifier.
\end{itemize}
all of which share the same LLM backbone as us. Recall@1 results are exhibited in Figure \ref{fig:snapshot2} (left), which implies:
\begin{itemize}
    \item {\ModelName} apparently outperforms the zero-shot Classifier, which lacks the domain-specific knowledge.
    \item Prompting tricks such as ICL, CoT, or Refine, can improve the Classifier's performance slightly; however, they are still worse than {\ModelName}.
    \item Finetuning can enhance the retrieval performance significantly, which is close to the Retriever and Reranker methods. However, our {\ModelName} is still better than it, revealing that the LLM may fail to capture the key information given the long prompt with all document contents included, while the retrieval-specific learning is more effective in this experiment.
\end{itemize}
From the aspect of inference latency, Figure \ref{fig:snapshot2} (right) further indicates that Classifier has a similar computation overhead as Reranker, both of which need to jointly process both query and document online. Multi-hop methods have even higher latency since they usually need to infer the LLM more than once. In contrast, our {\ModelName} has a similar speed to Retriever, indicating a good balance between performance and efficiency.

\subsection{Sensitivity}

\begin{figure}[t!]
\centering
  \includegraphics[width=0.48\linewidth]{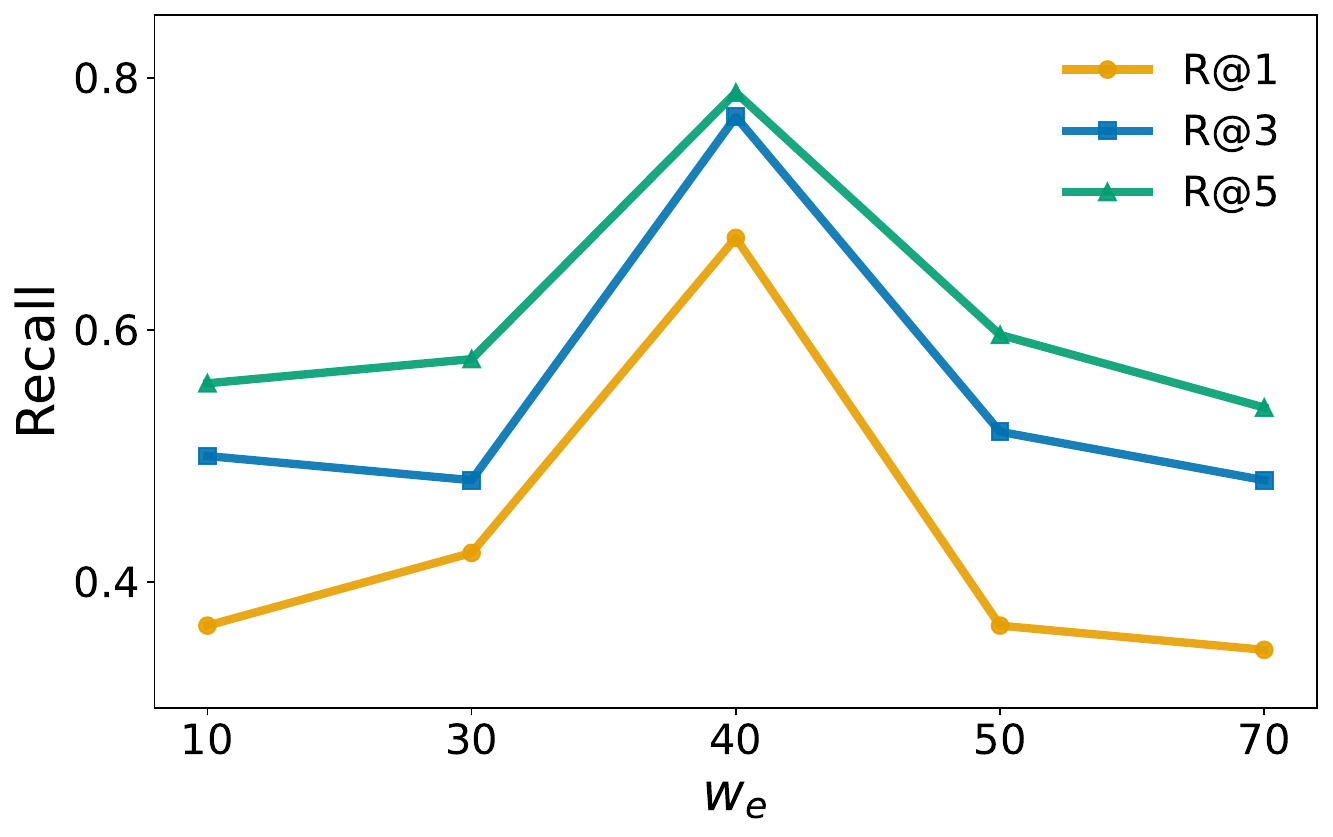}
  \hspace{0.05in}
  \includegraphics[width=0.48\linewidth]{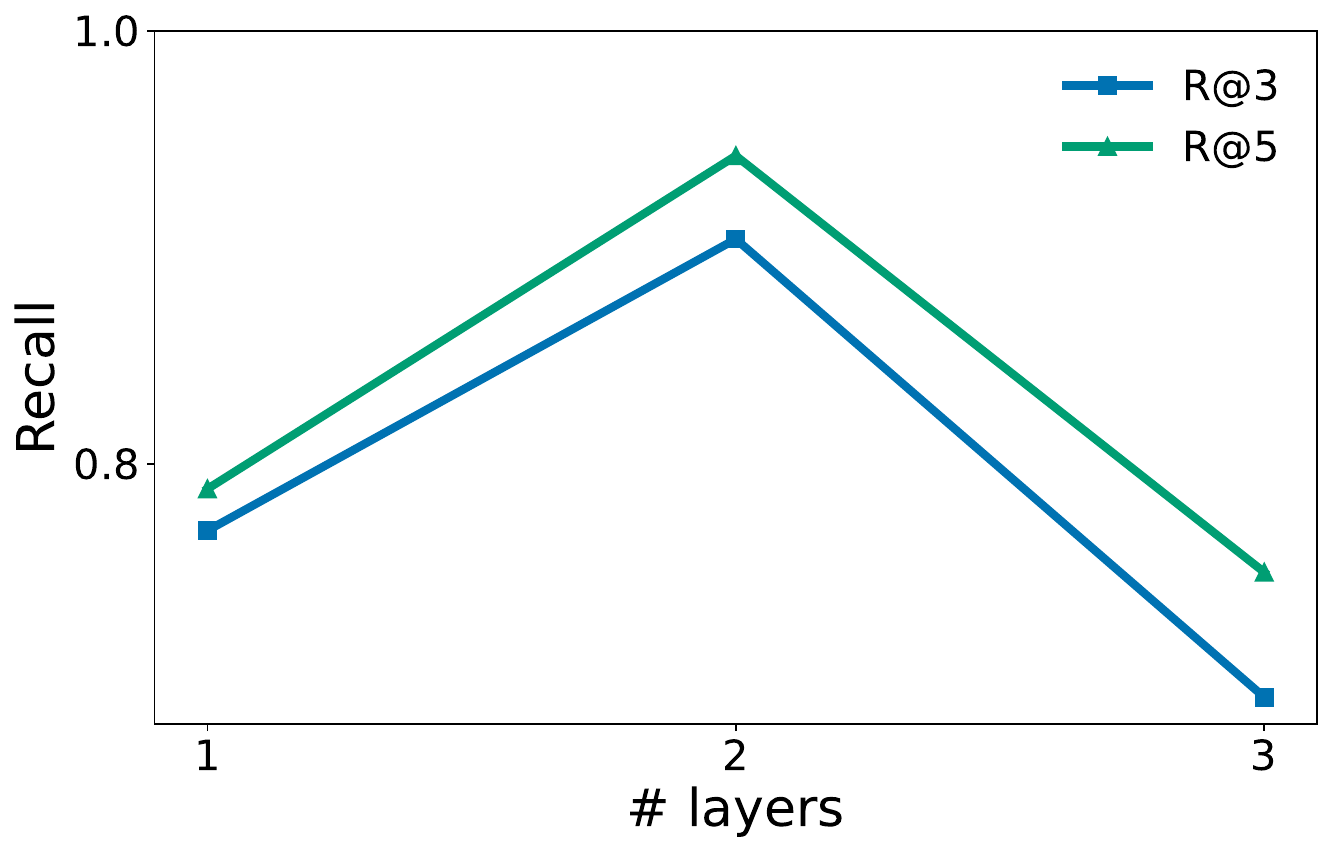}
  \caption{Recall curves of SayCan on varied loss weights $w_e$ (left) and the number of MLP layers (right).}
  \label{fig:sensitivity1}
\end{figure}

We test two critical hyperparameters, including the weight of the embedding loss $w_e$ (which equals the inverse of $w$ in Eq. (\ref{eq:student_objective})), and the number of MLP layers in the predictor. Based on Figure \ref{fig:sensitivity1}, it is evident that $w_e=40$ and $\# layer = 2$ performs the best, which is adopted in the formal experimental setting.

\subsection{End-to-End RAG}
We deploy {\ModelName} on an end-to-end RAG pipeline, with {\ModelName} behaving as the strategy retriever, and Qwen2-70B-Instruct as the response generator. We evaluate this pipeline on the test set of \textit{ESConv} by LLM-as-a-Judge based on GPT4. Pairwise evaluations are conducted between our response and the original responses in the datasets. {\ModelName} finally has a 80.12\% win-rate.


\subsection{Statistical Significance of Evaluation}
\label{sec:significant_test}

We assess the statistical significance by a two-sample t-test on R@1. For each method, we resample its responses 10 times, and calculate the standard deviations and corresponding p-values. Table \ref{tab:significant_test_ESConv} exhibits these results on ESConv. For both \textbf{teacher} and \textbf{student}, the reported p-values are calculated against LlaMA2Vec, the best retriever baseline. They are much below the $0.001$ significance level, indicating that the performance improvements of our {\ModelName} and \textbf{teacher} relative to LlaMA2Vec are highly statistically significant.

\begin{table}[htbp]
\centering
\small
\begin{tabular}{lccc}
\toprule
\textbf{Method} & \textbf{R@1} & \textbf{Std} &\textbf{p-value} \\
\midrule
LlaMA2Vec  & 23.9 & 0.023 & - \\
LaHoRe (\textbf{teacher}) & 34.4 & 0.038 & $< 10^{-5}$\\
\textbf{\ModelName} (ours) & 32.8 & 0.044 & $< 10^{-5}$\\
\bottomrule
\end{tabular}
\caption{R@1 results and statistical significances on ESConv. Significant tests are against LlaMA2Vec.}
\label{tab:significant_test_ESConv}
\end{table}

\section{Limitations}

While {\ModelName} offers significant gains in efficiency, its performance is inherently constrained by the quality of the cross-encoder teacher used for the JEPA-based distillation process. Additionally, although it is substantially faster than cross-encoders, the integration of a predictor block to model latent variables introduces a modest computational cost compared to traditional, interaction-free bi-encoder architectures. Lastly, as our experiments primarily focused on specific rationale-based tasks such as conversational strategies and robotic skills using relatively short document segments, the framework's scalability to broader, massive-scale general retrieval tasks remains to be fully explored.

\end{document}